\documentstyle[sprocl,epsf]{article}

\def\none               {\multicolumn{2}{c|}{---}}
\def\be{\begin{equation}}
\def\ee{\end{equation}}
\newcommand{\chinfty}{\chi^\infty}
\newcommand{\chiqu}{\chi^{\rm qu}}

\def\lsi{\raise0.3ex\hbox{$<$\kern-0.75em\raise-1.1ex\hbox{$\sim$}}}
\def\gsi{\raise0.3ex\hbox{$>$\kern-0.75em\raise-1.1ex\hbox{$\sim$}}}

\newcommand{\MeV}{\mathop{\rm MeV}}
\newcommand{\fm}{\mathop{\rm fm}}

\begin{document}
\title{
\vspace*{-8ex}
{\protect \footnotesize\rm \flushright
Edinburgh 2000/19 \\
hep-lat/0009008 \\
September 2000 \\}
\vspace{2ex}
THE TOPOLOGICAL SUSCEPTIBILITY AND PION  \\
DECAY CONSTANT FROM LATTICE QCD}

\author{{\sl UKQCD Collaboration:} A. HART%
\footnote{Talk presented at the Confinement IV meeting, Vienna, July 2000.}}

\address{Dept. of Physics and Astronomy, Univ. of Edinburgh,
Edinburgh, Scotland}

\author{and M. TEPER}

\address{Theoretical Physics, Univ. of Oxford, 1 Keble Road, 
Oxford, England}

\maketitle\abstracts{ We study the topological susceptibility, $\chi$,
  in two flavour lattice QCD. 
\cite{hart_other}
We find clear evidence for the expected suppression of $\chi$ at small
quark mass. The estimate of the pion decay constant, $f_{\pi} = 105
\pm 5 \ ^{+18}_{-10} \ \MeV$, is consistent with the experimental
value of approximately $93 \ \MeV$.  We compare $\chi$ to the
large-$N_c$ prediction and find consistency over a large range of
quark masses.}

The ability to access the non--perturbative sectors, and to vary
parameters fixed in Nature has made lattice Monte Carlo simulation a
valuable tool for investigating the r\^{o}le of topological excitations
in QCD and related theories.
\cite{teper99}

The topological susceptibility is the squared expectation value of the
topological charge, normalised by the volume
\be
\chi = \frac{\langle Q^2 \rangle}{V},
\hbox{\hspace{2em}}
Q = \frac{1}{32\pi^2} \int d^4x 
\frac{1}{2} \varepsilon_{\mu \nu \sigma \tau}
F^a_{\mu \nu}(x) F^a_{\sigma \tau}(x).
\ee
Sea quarks induce an instanton--anti-instanton attraction which
in the chiral limit becomes stronger, suppressing $Q$ and $\chi$
\cite{vecchia80}
\be
\chi = \Sigma \left( {m_{u}}^{-1} + {m_{d}}^{-1} 
 \right)^{-1},
\hbox{\hspace{1em} where \hspace{1em}} 
\Sigma = - \lim_{m_q \to 0} \lim_{V \to \infty} 
\langle 0 | \bar{\psi} \psi | 0 \rangle
\ee
is the chiral condensate.
\cite{leutwyler92} 
We assume
$\langle 0 | \bar{\psi} \psi | 0 \rangle =
\langle 0 | \bar{u} u | 0 \rangle =
\langle 0 | \bar{d} d | 0 \rangle $
and neglect contributions of heavier quarks.
The  Gell-Mann--Oakes--Renner relation,
\be
f_\pi^2 m_\pi^2 = (m_{u} + m_{d}) \Sigma + {\cal O}(m_q^2)
\hbox{\hspace{1em}}
\Rightarrow
\hbox{\hspace{1em}}
\chi = \frac{f_\pi^2 m_\pi^2}{2 N_f} + {\cal O}(m_\pi^4)
\label{eqn_chi_pi2}
\ee
for $N_f$ light flavours, in a convention where the experimental value
of the pion decay constant $f_\pi \simeq 93 \ \MeV$.
Eq.~\ref{eqn_chi_pi2} holds in the limit $f_\pi^2 m_\pi^2 V \gg 1$,
which is satisfied by all our lattices.  The higher order terms ensure
that $\chi \rightarrow \chiqu$, the quenched value, as $m_q,m_\pi \to
\infty$.  We find, however, that our measured values are not very much
smaller than $\chiqu$, so we must consider two possibilities.

\begin{figure}[t]
\caption[]{\label{fig_chi}
  { The measured topological susceptibility, with interpolated
    quenched points at the same $\hat{r}_0$ and fits independent of
    the quenched points.}}
\vspace*{-30ex}
\leavevmode
\hspace*{2em}
\epsfysize=260pt
\epsfbox[20 30 620 730]{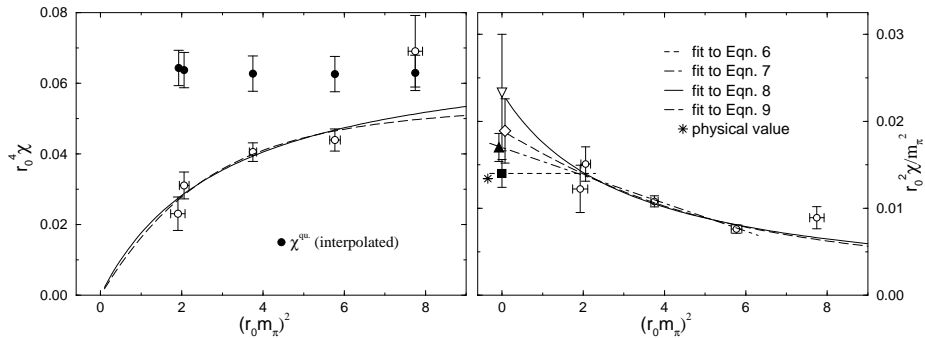}
\vspace{-2ex}
\end{figure}
Firstly, there are phenomenological reasons
\cite{thooft74,witten79}
for believing that QCD is `close' to $N_c = \infty$, and in the case
of gluodynamics even SU(2) is demonstrably close to SU($\infty$).
\cite{teper98,teper99}
Fermion effects are non-leading in $N_c$, so we expect $\chi \to
\chiqu$ for any fixed value of $m_q$ as the number of colours $N_c
\rightarrow \infty$. For small $m_q$ we expect Eq.~4 to hold, with
$\chinfty$, $f_\infty$ the quantities at leading order in $N_c$.
\cite{leutwyler92} 
Alternatively, our $m_q \simeq m_{\rm strange}$ and perhaps
higher order terms are important. In the
absence of a QCD prediction, Eq.~5 interpolates
between Eq.~\ref{eqn_chi_pi2} and the quenched limit.
$$
\chi = \frac{\chinfty m_\pi^2}
{\frac{2 N_f \chinfty}{f_\infty^2} + m_\pi^2}
\ \ (4), \mbox{\hspace{2em}}
\chi  =  
\frac{f_{\pi}^2}{\pi N_f}  m_{\pi}^2
\arctan \left(\frac{\pi N_f}{f_{\pi}^2} \chiqu
\frac{1}{m_{\pi}^2} \right)
\ \ (5).
$$
Measurements of $\chi$ were made on a number of ensembles of $N_f=2$
lattice field configurations produced by the UKQCD Collaboration. An
SU(3) Wilson gauge action is coupled to clover improved Wilson
fermions.
\cite{ukqcd99}
The UKQCD ensembles have two notable features. The improvement is
fully non--perturbative, with discretisation errors being quadratic
rather than linear in the lattice spacing. Second, the couplings are
chosen to maintain an approximately constant lattice spacing (as
defined by the Sommer scale, $r_0=0.49 \ \fm$~%
\cite{sommer94}) 
as the quark mass is varied. This is important, as the susceptibility
in gluodynamics varies with the lattice spacing as
\cite{teper99}
$
\hat{r}_0^4 \hat{\chi} = 0.072 - 0.208/\hat{r}_0^2
$ 
in competition with the variation with $m_q$.
\cite{hart_other}
The topological susceptibility is measured from the gauge fields after
cooling to remove the UV noise. We plot these data in
Fig.~\ref{fig_chi} along with the interpolated $\chiqu$ at an
equivalent lattice spacing from the above formula for comparison,
which vary little owing to the the UKQCD matching.  The behaviour with
$M \equiv (\hat{r_0} \hat{m}_\pi)^2$ is qualitatively as expected and,
more quantitatively, we attempt fits motivated by
Eqs.~\ref{eqn_chi_pi2},~4,~5:
$$
\frac{\hat{r_0}^2 \hat{\chi}}{M}=
c_0 \ \ (6), 
c_0 + c_1 (\hat{r_0} \hat{m}_\pi)^2 \ \ (7),
\frac{c_0 c_3}
{c_3 + c_0 M} \ \ (8),
\frac{2c_0}{\pi} 
\tan^{-1} \left( \frac{\pi c_3}{2 c_0 M} \right) \ \ (9)
\nonumber
$$
\begin{table}[t]
\caption{ \label{tab_fit_chi}
  { Fits to the $N_{\rm fit}$ most chiral points of
    $\hat{\chi}$.}}
\vspace{0.2cm}
\begin{center}
\footnotesize
\begin{tabular}{|c|c|r@{.}l|r@{.}l|r@{.}l|r@{.}l|}
\hline
Fit & 
$N_{\rm fit}$ &
\multicolumn{2}{c|}{$c_0$} &
\multicolumn{2}{c|}{$c_1$} &
\multicolumn{2}{c|}{$\chi^2/{\rm d.o.f.}$} &
\multicolumn{2}{c|}{$\hat{r}_0 \hat{f}_\pi$} \\
\hline
Eq.~6 & 2 &  
0&0140 (16)  & \none & 0&805 &  0&237 (14) \\
Eq.~6 & 3 &  
0&0112 (6) & \none & 2&202 &  0&212 (6)  \\
Eq.~6 & 4 &  
0&0091 (4) & \none & 9&008 & \none \\
\hline
Eq.~7 & 3  & 
0&0176 (35) (4)  & $-0$&0018 (10) (1) & 0&964 & 0&265 (27) \\
Eq.~7 & 4  & 
0&0170 (16) (1)  & $-0$&0016 (4) (0) & 0&502  & 0&261 (13) \\
Eq.~7 & 5  & 
0&0147 (14) (1)  & $-0$&0011 (3) (0) & 2&965 & 0&242 (12) \\
\hline
\hline
Fit & 
$N_{\rm fit}$ &
\multicolumn{2}{c|}{$c_0$} &
\multicolumn{2}{c|}{$c_3$} &
\multicolumn{2}{c|}{$\chi^2/{\rm d.o.f.}$} &
\multicolumn{2}{c|}{$\hat{r}_0 \hat{f}_\pi$} \\
\hline
Eq.~8 & 3 & 
0&0208 (87) (12) & 0&0844 (427) (35) &  
1&013 & 0&288 (61) \\
Eq.~8 & 4 & 
0&0272 (85) (18) & 0&0632 (114) (6) &  
0&895 & 0&329 (53) \\
Eq.~8 & 5 & 
0&0233 (66) (10) & 0&0717 (147) (3) &  
1&847 &  0&305 (44) \\
\hline
Eq.~9 & 3 & 
0&0186 (53) (7) & 0&0576 (175) (6) &  
0&990 & 0&273 (40) \\
Eq.~9 & 4 & 
0&0209 (42) (7) & 0&0506 (55) (5) &  
0&682 & 0&289 (30) \\
Eq.~9 & 5 & 
0&0189 (36) (5) & 0&0550 (69) (6) &  
1&929 &  0&275 (27) \\
\hline
\end{tabular}
\end{center}
\vspace{-2ex}
\end{table}
\noindent
We include progressively less chiral points until the fit becomes
unacceptably bad in Table~\ref{tab_fit_chi}. We note the wide range
fitted simply by including an $m_\pi^4$ term, and the consistency of
our data with large-$N_c$ predictions. The stability and similarity of
the fits motivates us to use $c_0$ from Eq.~7 to estimate
$f_\pi = 105 \ \pm 5 \ ^{+18}_{-10} \ \MeV$, with variation between
other fits providing the second, systematic error, and in good
agreement with the experimental value $\simeq 93 \ \MeV$.

\vspace{-2ex}
\section*{References}

{%

}

\end{document}